\PassOptionsToPackage{pdfpagelabels=false}{hyperref} 
\documentclass[usenatbib,fleqn]{mnras}
\usepackage{fixltx2e,ifpdf,amsmath,amssymb,fancyhdr,nicefrac,wasysym,marvosym}
\usepackage{times,nicefrac,bm}
\usepackage[multiple]{footmisc}
\ifpdf
  \usepackage{aeguill}
  \usepackage[pdftex]{graphicx,color}
\else
  \usepackage[dvips]{graphicx,color}
\fi
\newcommand{\rpole} {\ensuremath{R_{{\rm p}*}}}
\newcommand{\reqtr} {\ensuremath{R_{{\rm e}}}}
\newcommand{\tpole} {\ensuremath{T_{{\rm p}}}}
\newcommand{\teqtr} {\ensuremath{T_{{\rm e}}}}
\newcommand{\porb} {\ensuremath{P_{\rm orb}}}
\newcommand{\prot} {\ensuremath{P_{\rm rot}}}

\newcommand{\istar} {\ensuremath{i_*}}
\newcommand{\iorb} {\ensuremath{i_{\rm orb}}}
\newcommand{\veq}   {\ensuremath{v_{\rm e}}}
\newcommand{\vesini}{\ensuremath{v_{\rm e}\sin{i_*}}}
\newcommand{\omomc}{\ensuremath{ \Omega/\Omega_{\rm c} }}
\newcommand{\OmOmc}{\ensuremath{ \frac{\Omega}{\Omega_{\rm c}} }}
\newcommand{\teff}{\ensuremath{T_{\rm eff}}}
\newcommand{\teffl}{\ensuremath{T^{\ell}_{\rm eff}}}
\newcommand{\teffL}{\ensuremath{T^{\rm L}_{\rm eff}}}
\newcommand{\logg}{\ensuremath{\log{g}}}
\newcommand{\loggp}{\ensuremath{\log{g}_{\rm p}}}
\newcommand{\loggpL}{\ensuremath{\log{g}^{\rm L}_{\rm p}}}
\newcommand{\kms} {\ensuremath{\mbox{km}\;\mbox{s}^{-1}}}
\newcommand{\msun}{\ensuremath{\mbox{M}_{\odot}}}
\newcommand{\rsun}{\ensuremath{\mbox{R}_{\odot}}}
\newcommand{\rjup}{\ensuremath{\mbox{R}_{\jupiter}}}
\newcommand{\lsun}{\ensuremath{\mbox{L}_{\odot}}}
\newcommand{\mstar}{\ensuremath{M_*}}
\newcommand{\rstar}{\ensuremath{R_*}}
\newcommand{\pzero}{\ensuremath{\phantom{0}}}
\newcommand{\cdfil}{\xleaders\hbox{{---}}\hfill\kern0pt}

\def\Vhrulefill{\leavevmode\leaders\hrule height 0.7ex depth \dimexpr0.4pt-0.7ex\hfill\kern0pt}

\title[Rapid rotators revisited:  KOI-13] 
{Rapid rotators revisited:  absolute dimensions of KOI-13}
\author[I.\ D.\ Howarth and G.\ Morello]
{Ian D.\ Howarth\thanks{E-mail: i.howarth@ucl.ac.uk} and Giuseppe Morello\\
  Dept.\ of Physics and Astronomy, University College London, Gower
  Street, London WC1E 6BT, UK }
\begin{document}

\date{Accepted 2017 May 17. Received 2017 May 14; in original form
  2017 April 19}

%\pagerange{\pageref{firstpage}--\pageref{lastpage}} \pubyear{2012}

\maketitle

\label{firstpage}

% -------------------------------------------------------------------------------------------------
% Abstract
\begin{abstract}
We analyse \textit{Kepler} light-curves of the exo\-planet KOI-13b
transiting its moderately rapidly rotating (gravity-darkened) parent
star.  A physical model, with minimal ad hoc free parameters,
reproduces the time-averaged light-curve at the $\sim$10 parts per
million level.  We demonstrate that this Roche-model solution allows
the absolute dimensions of the system to be determined from the
star's projected equatorial rotation speed, \vesini, without any
additional assumptions;  we find 
a planetary radius 
\mbox{$R_{\rm P} = (1.33 \pm 0.05)~\rjup$,}
stellar polar radius 
$\rpole = (1.55 \pm 0.06)~\rsun$,
combined mass 
$\mstar + M_{\rm P} (\simeq{\mstar}) = (1.47 \pm 0.17)~\msun$,
and distance $d \simeq (370 \pm 25)$~pc, where the errors are dominated
by uncertainties in relative flux contribution of the visual-binary
companion KOI-13B.  The implied stellar rotation period is within $\sim$5\%\ of
the non-orbital, 25.43-hr signal found in the \textit{Kepler} photometry.  We
show that the model accurately reproduces independent tomo\-graphic
observations, and yields an offset between orbital and
stellar-rotation angular-momentum vectors of $60{\fdg}25\pm0{\fdg}05$.

\end{abstract}


% -------------------------------------------------------------------------------------------------
% Introduction
\section{Introduction}

One of the many unexpected results to emerge from studies of
exo\-planets this century has been the discovery of orbits
that are not even approximately coplanar with the stellar equator
(cf., e.g., \citealt{Winn15}).

The tool traditionally most commonly used to investigate the
relative orientations of orbital and stellar-rotation angular-momentum
vectors is the Rossiter--McLaughlin (\mbox{R--M}) effect
(\citealt{Holt1893, Schlesinger1910}\footnote{An example of Stigler's
law \citep{Merton57,Stigler80}.}) -- the apparent displacement of rotationally broadened stellar
line profiles arising from a body occulting part of the stellar disk.
Long established in eclipsing-binary
studies \citep[e.g.,][]{Rossiter24, McLaughlin24}, the R--M effect
took on new significance following its detection in the archetypal
transiting exo\-planetary system HD~209458 \citep{Queloz00}.  The
discovery of misaligned planetary orbits in other systems followed \citep{Hebrard08,
Winn09}, and sample sizes are now large enough\footnote{$\sim$120 at
the time of writing;
e.g.,\newline \texttt{http://www.astro.keele.ac.uk/jkt/tepcat/rossiter.html}}
to suggest that stars with thick convective envelopes generally have planets
with small orbital misalignments, while a broader spread of
values is found in hotter stars
\citep{Winn10,Schlaufman10,Albrecht12,Mazeh15}.

%Other techniques for investigating spin-orbit alignment include
%asteroseismology (e.g., \citealt{Chaplin13, Huber13,Campante16}),
%astrometry \citep{leBouquin09}

The R--M effect is an essentially spectro\-scopic phenomenon, being studied through
radial-velocity measurements.   
In principle there is a corresponding
photo\-metric signature, arising through Doppler boosting (e.g.,
\citealt{Groot12}), but the signal is too small for any reliable
detections to date.  Transit photo\-metry does, however, offer
potential diagnostics of spin-orbit alignment if the
surface-brightness distribution over the occulted parts of the stellar
disk is not circularly symmetric.  In particular, if the stellar
rotation is sufficiently rapid, it can introduce both an equatorial
extension and, through gravity darkening, a characteristic
latitude-dependent surface-intensity distribution; these effects are
capable of defining the relative direction of the stellar rotation
axis, and hence of diagnosing misaligned transits (e.g., \citealt{Barnes09}).

The first system to be recognized as having a misaligned orbit from
photo\-metry alone, without supporting evidence from the R--M effect,
was KOI-13 \citep{Szabo11,Barnes11}.  Other systems in which asymmetry
in the transit light-curve has been interpreted as arising through
rotationally-induced gravity darkening include KOI-89 \citep{Ahlers15}
and HAT-P-7 (KOI-2; \citealt{Masuda15}),  while the same approach has
been used to argue for good alignment of orbital and rotational
angular-momentum vectors for KOI-2138 \citep{Barnes15}.
In other cases,
  modelling of lower-quality data has led to less compelling claims;
  e.g., PTFO~8-8695 (cp.\ \citealt{Barnes13, Howarth16}) and CoRot-29
  (cp.\ \citealt{Cabrera15, Palle16}).

In the present paper we re-examine \textit{Kepler} photo\-metry of transits of
KOI-13, using a more complete physical model than previous studies.
Our intention is to stress-test the model against data of remarkable
quality, and to demonstrate its power to  establish \textit{absolute}
numerical values for key stellar and planetary parameters.
Following a selective review of the literature on KOI-13 
($\S$\ref{sec:koi}), we summarize the model ($\S$\ref{sec:mod}) and the
data preparation ($\S$\ref{sec:dprep}).   Results are presented and
discussed in $\S\S$\ref{sec:fit},~\ref{sec:syspar}.   Appendix~\ref{appx:one}
demonstrates how to put the modelling on an absolute scale,
given the star's projected equatorial rotation speed.

\section{The KOI-13 system}
\label{sec:koi}

Kepler Object of Interest no.\ 13 (KOI-13; historically cata\-logued as
BD~+46$^\circ$~2629) was identified as the host of a transiting
exo\-planet by \citet{Borucki11}.  \citet{Aitken04} had previously noted
BD~+46$^\circ$~2629 as visual binary with components of comparable
brightness, separated by $\sim$1{\farcs}1 \citep{Howell11, Law14},
which \citet{Szabo11} showed share a common proper motion.  The latter
authors identified the marginally brighter component as the transiting
system, a result confirmed by \citet{Santerne12}, who found
the fainter component, KOI-13B, to be itself a spectro\-scopic binary.

The basic transit light-curve was modelled
by \citet{Barnes11}, who showed that its small asymmetry 
arises from stellar gravity darkening coupled to
spin--orbit misalignment.  Subsequent tomo\-graphy yielded results
inconsistent with the obliquity inferred in this first analysis
\citep{Johnson14}, but by imposing the constraint afforded by the
spectroscopy,
\citet{Masuda15} was able to identify a geometry that reconciled
the spectro\-scopic and light-curve solutions.

The exquisite quality of the \textit{Kepler} data has inspired a number of
ancillary studies.  In particular, the system clearly shows
out-of-transit orbital variations arising from Doppler beaming,
ellipsoidal distortion, and reflection effects (\textsc{`beer'}
effects; \citealt{Shporer11, Mislis12, Mazeh12}).  A further,
25.43-hr, periodic signal has been identified in the photometry, and
has been suggested as arising either from tidally induced pulsation
\citep{Shporer11,Mazeh12} or from rotational
modulation \citep{Szabo12}.

\section{Modelling}
\label{sec:mod}

The \citet{Barnes11} and \citet{Masuda15} analyses 
of the transit light-curve
were both based on
a simple oblate-spheroid stellar geometry, and utilised black-body
fluxes coupled to a global two-parameter limb-darkening `law'.
These are reasonable approximations for initial investigations, especially since KOI-13's
rotation is substantially subcritical (cf.\ Table~\ref{params_tab}), but we undertook
our work in the hope that a somewhat more physically-based model would
better constrain the system with fewer ad hoc adjustments.

The basic model is as described by \citeauthor{Howarth16}
(\citeyear{Howarth16}; \citealt{Howarth01}).  Appropriate values for
model parameters, and their probability distributions, are determined
through Markov-chain Monte-Carlo (MCMC) sampling, with uniform priors
unless stated otherwise.

\subsection{Star}

The star's rotationally distorted surface is approximated as a Roche
equipotential.\footnote{Mass distributions from polytropic models give
negligibly different results \citep{Plavec58, Martin70}.  By default,
surface angular velocity is assumed to be independent of latitude.}
Latitude-dependent values of surface gravity, $g$,
and \textit{local} effective temperature, \teffl, are calculated
self-consistently, taking into account gravity darkening.  
The stellar flux is then computed as a
numerical integration of emitted intensities over visible surface elements.

\subsubsection{Intensities}

Specific intensities (radiances), $I(\lambda, \mu, \teffl, g)$,
are interpolated from a grid of line-blanketed, solar-abundance LTE
models \citep{Howarth11a}, integrated over the \textit{Kepler}
passband.  
The interpolation in angle ($\mu =\cos\theta$, where $\theta$
is the angle between the surface normal and the line of sight)
is performed using an analytical
4-parameter characterization
\begin{align}
I(\mu)/I(1) = 1 - \sum_{n=1}^4{a_n (1 - \mu^{n/2})}
\label{eq:cl4}
\end{align}
\citep{Claret00}, which reproduces individual numerical values to
$\sim$0.1\%\ \citep{Howarth11a}. 

\subsubsection{Modelled effective temperature, gravity}
\label{sec:teffl}

Surface distributions of temperature and gravity are needed in order
to evaluate model-atmosphere emergent intensities (and for no other
reason).  These parameters are completely specified by the adopted
gravity-darkening law ($\S$\ref{sec:gd}), plus any suitable
normalizations; we use the base-10 logarithm of the polar gravity in
c.g.s. units, \loggp, and the stellar effective temperature,
\begin{align*}
\teff = \sqrt[4]{\frac{\int{\sigma(\teffl)^4\,\text{d}A}}{{\int{\sigma\,\text{d}A}}}}
\end{align*}
(where $\sigma$ is the Stefan--Boltzmann constant and the integrations
are over surface area).

While the use of model-atmosphere intensities removes the need for ad
hoc limb-darkening parameters, this is at the expense of assumptions
that, first, the effective temperature and polar gravity are known
with adequate precision to give a sufficiently faithful representation of limb
darkening, and secondly, that the model-atmosphere calculations
predict the emergent intensities reliably.
Anticipating that
neither assumption need necessarily be valid (e.g., \citealt{Howarth11b}),
we draw an explicit distinction between the actual physical
quantities \teff, \loggp\ and their model-parameter counterparts
\teffL, \loggpL\ (where the superscript 
is intended to indicate a `light-curve', or `limb-darkening',
determination;  cf.~$\S$\ref{sec:fit}).

\subsubsection{Gravity darkening}
\label{sec:gd}

It is not immediately obvious whether gravity darkening in KOI-13
should be modelled according to a recipe appropriate for radiative or
convective envelopes.  While the literature documents a surprising
large dispersion for estimates of its effective temperature
(7650--9107~K; \citealt{Shporer14}, \citealt{Szabo11},
\citealt{Brown11}, \citealt{Huber14}, with claimed precisions that are
considerably smaller than the spread of results), the more detailed
studies tend towards values at the lower end of the range.  This puts
\teff\ not very far from the boundary between convective and radiative
regimes, around $\teff \simeq 7000$~K (e.g.,
\citealt{Claret98}).  Because of this, we ran several sequences of
models using a generic gravity-darkening law,
\begin{align}
\teffl \propto g^\beta,
\label{eq:gdark}
\end{align}
with the
gravity-darkening exponent $\beta$ as a free parameter.  These models
all migrated to solutions with exponents very close to the
\citet{vonZeipel24}
value
of $\beta=0.25$, as was also found by
\citet{Masuda15}.  

For most model runs, we actually used the parameter-free
gravity-darkening model proposed by \citet{Espinosa11}, which is close
to von~Zeipel gravity darkening at the subcritical rotation
appropriate to KOI-13.
This `ELR' formulation has a somewhat firmer physical foundation than
the original von~Zeipel analysis, and gives better agreement with, in
particular, optical interferometry of rapid rotators
(e.g., \citealt{DomdeSou14}).

\subsection{Transit}

Transits are modelled by  assuming a completely dark occulting body of
circular cross-section, in a misaligned circular orbit;
although an orbital eccentricity $e = (6\pm 1) \times 10^{-4}$ has
been inferred from out-of-transit photo\-metry of KOI-13 by
\citet{Esteves15}, this has negligible consequences for our study.
The contamination of the transit light-curve by KOI-13B (spatially
unresolved in the \textit{Kepler} beam) is characterized by its
fractional contribution to the total signal, or `third light' ($L_3$)
in the nomenclature of traditional eclipsing-binary
studies.\footnote{Of course, the exo\-planetary `second light' is
extremely small.}

\subsection{Parameters}

Table~\ref{params_tab} lists one set of basic parameters that fully
specify the model (other combinations are possible).  We stress that
the geometry of the model is fundamentally scale-free; all linear
dimensions are expressed in units of the orbital semi-major axis,
while times are implicitly in units of the orbital period.  The extent
of effects arising from rotational distortion is determined by
$\Omega/\Omega_{\rm c}$, the ratio of the rotational angular velocity
to the critical value at which the effective equatorial gravity is
zero; a value for the stellar mass, often assumed in similar studies,
is not required.

\begin{table*}
\caption{Model parameters and illustrative fitted values.  Model M1 has \teffL\ as
a free parameter (cf.~$\S$\ref{sec:teffl}), with $\loggpL \equiv \loggp$;  model M2 additionally has 
\loggpL\ free;  model M3 has \prot\ fixed.
The errors (on the last quoted significant figure of the parameter
values) are the quadratic sum of 95-percentile ranges on solution M1
(initial $L_3 = 0.45$) and the maximum deviation of corresponding
solutions with initial $L_3$ values in the range 0.41--0.49
($\S$\ref{sec:l3}).}
\begin{tabular}{llllrlll}
\hline
\multicolumn{3}{c}{Parameter} &
\multicolumn{5}{c}{{\Vhrulefill}\;Best-fit value\;\Vhrulefill} 
\\
&&\multicolumn{1}{r}{Model:}&\multicolumn{1}{c}{M1} 
&\multicolumn{1}{c}{$\pm$} &$\qquad$&
\multicolumn{1}{c}{M2}&
\multicolumn{1}{c}{M3}\\
\hline

\multicolumn{3}{l}{{Stellar:}} \\
$\;$& \teffL & Effective-temperature parameter$^*$ (K)&
\pzero\ 8084& 186&&\pzero\ 7987&$\phantom{\equiv}$\pzero\ 8046\\
$\;$& \loggpL & Polar-gravity parameter$^*$ (dex cgs)&
\;\;\;$\cdots$& &&\pzero 4.27&$\phantom{\equiv}$\pzero 4.32\\
& \omomc  &
\begin{minipage}[t]{0.7\columnwidth}
Angular rotation rate\newline (in units of the critical rate)
\end{minipage}&
\pzero 0.341& 15&&\pzero 0.343&$\phantom{\equiv}$\pzero 0.320\\
& $\istar$ & 
\begin{minipage}[t]{0.7\columnwidth}
Inclination of stellar rotation axis to line of sight \mbox{(0--90$^\circ$)}
\end{minipage}&
81.137&16&&81.135&$\phantom{\equiv}$81.134\\
&$\rpole/a$& 
\begin{minipage}[t]{0.7\columnwidth}
Polar radius\newline(in units of the orbital semi-major axis)
\end{minipage}&
\pzero 0.2219&4&&0.2217&$\phantom{\equiv}$\pzero 0.2219\\
&$L_3$&`Third light'&
\pzero 0.451&39&&\pzero 0.451&$\phantom{\equiv}$\pzero 0.451\\
&g.d.& 
\begin{minipage}[t]{0.7\columnwidth}
Gravity darkening: ELR
\end{minipage}&
\\

\multicolumn{3}{l}{{Planetary:}} \\
&$R_{\rm P}/a$ & 
\begin{minipage}[t]{0.7\columnwidth}
Planetary radius\\(in units of the orbital semi-major axis)
\end{minipage}&
\pzero 0.0190&7&&\pzero 0.0190&$\phantom{\equiv}$\pzero 0.0190\\
\multicolumn{3}{l}{{Orbital:}} \\
&\iorb & 
\begin{minipage}[t]{0.7\columnwidth}
Inclination of orbital angular-momentum vector to line
of sight \mbox{(0--180$^\circ$)}
\end{minipage}&
93.319&22&&93.316&$\phantom{\equiv}$93.316\\
&$\lambda$ &
\begin{minipage}[t]{0.7\columnwidth}
Angle between the projections onto the plane of the sky
of the orbital and stellar-rotational angular-momentum vectors, 
measured counter-clockwise from the former \mbox{(0--360$^\circ$)}
\end{minipage}
&59.19&5&&59.20&$\phantom{\equiv}$59.20\\
%
%&$T_0$ & 
%\begin{minipage}[t]{0.7\columnwidth}
%Time of conjunction (HJD)
%\end{minipage}\\
\hline
\multicolumn{3}{l}{Imposed:} \\
&$\porb$ & 
\begin{minipage}[t]{0.7\columnwidth}
Orbital period (d)
\end{minipage}&\multicolumn{5}{c}{{\Vhrulefill}{\;1.76358799\;}{\Vhrulefill}}\\

& \vesini & Projected equatorial rotation speed$^\dagger$ (\kms)&
\multicolumn{5}{c}{{\Vhrulefill}\;$76.6 \pm 0.2$\;{\Vhrulefill}}\\
&\prot&Rotation period (d)&
$\;\;\;\cdots$&&&$\;\;\;\cdots$&$\phantom{\equiv}$\pzero 1.0596\\

\multicolumn{3}{l}{Derived stellar parameters:} \\
& \loggp & True polar gravity (dex cgs)&
\pzero 4.209&19&&\pzero 4.21&$\phantom{\equiv}$\pzero 4.24\\
&\rpole/\rsun& Polar radius&
\pzero 1.49&7&&\pzero 1.48&$\phantom{\equiv}$\pzero 1.61\\
&\reqtr/\rsun&Equatorial radius&
\pzero 1.52&7&&\pzero 1.51&$\phantom{\equiv}$\pzero 1.63\\
&Oblateness&$1-\rpole/\reqtr$&
\pzero 0.0178&17&&\pzero 0.0181&$\phantom{\equiv}$\pzero 0.0156\\
&\tpole/\teff&Relative polar temperature&
\pzero 1.0118&11&&\pzero 1.0119&$\phantom{\equiv}$\pzero 1.0103\\
&\teqtr/\teff&Relative equatorial temperature&
\pzero 0.9939&6&&\pzero 0.9938&$\phantom{\equiv}$\pzero 0.9947\\
&$(1+q)\mstar/\msun$&System mass$^\ddagger$&
\pzero 1.31&17&&\pzero 1.29&$\phantom{\equiv}$\pzero 1.64\\
&$\log(L^{\rm L}/\lsun)$&$\text{luminosity} \times(\teff/\teffL)^4$
(dex solar)&
\pzero 0.94&3&&\pzero 0.92&$\phantom{\equiv}$\pzero 1.00\\
&$\rho_*$&Mean density (g~cm$^{-3}$)&
\pzero 0.5373&11&&\pzero 0.5380&$\phantom{\equiv}$\pzero 0.5397\\
&\veq&Equatorial rotation speed (\kms)&
77.3&4&&77.5&$\phantom{\equiv}$78.0\\
&\prot&Rotation period (d)&
\pzero 0.994&23&&\pzero 0.987&$\phantom{\equiv}\;\pzero \cdots$\\

\multicolumn{3}{l}{Other derived parameters:} \\
& $R_{\rm P}/\rjup$ & Planetary radius
$(\rjup = ({\mathcal{R}^{\rm N}_{e{\rm
J}}  \mathcal{R}^{\rm N}_{p{\rm J}}})^{1/2})$
&
\pzero 1.28&5&&\pzero 1.28&$\phantom{\equiv}$\pzero 1.38\\
&$\psi$& 
\begin{minipage}[t]{0.7\columnwidth}
Angle between orbital and stellar-rotational angular-momentum vectors (0--180$^\circ$)
\end{minipage}&
60.24&5&&60.25&$\phantom{\equiv}$60.25\\
&$b$&Impact parameter (\rpole)&
\pzero 0.2609&12&&\pzero 0.2609&$\phantom{\equiv}$\pzero 0.2607\\
\hline
\multicolumn{8}{l}{
\begin{minipage}[t]{0.86\textwidth}
%\scriptsize 
Additional model parameters include $e$,
the orbital eccentricity ($e=0$ assumed here)
and longitude of periastron (0--360$^\circ$;  undefined when $e=0$).
Best-fit (minimum-$\chi^2$) parameter sets are listed;  median values
of MCMC runs are extremely close to these values.
%\normalsize
\end{minipage}}\\
\multicolumn{8}{l}
{\begin{minipage}[t]{0.86\textwidth}
%\scriptsize
$^*$Used only to evaluate model-atmosphere intensities, and
constrained in the present study only by limb darkening; cf.\
$\S$\ref{sec:teff}\\
$^\dagger$Derived radii scale linearly with \vesini, and the mass as
$(\vesini)^3$;  Appendix~\ref{appx:one}.\\
$^\ddagger$Mass ratio $q \equiv M_{\rm P}/\mstar \simeq 4 \times
10^{-3}$ (\citealt{Shporer14};
\citealt{Esteves15};
\citealt{Faigler15}).\normalsize\end{minipage}}\\
%\multicolumn{8}{l}
%%\scriptsize
%{$^\ddagger$Mass ratio $q \equiv M_{\rm P}/\mstar \simeq 4 \times
%10^{-3}$ (\citealt{Shporer14};
%\citealt{Esteves15};
%\citealt{Faigler15}).}\\
%%

\end{tabular}
\label{params_tab}
\end{table*}

\section{Data preparation}
\label{sec:dprep}

We used the full set of short-cadence Pre-search Data
Conditioning Simple Aperture Photometry (PDCSAP) data, which are
publicly available through the \textit{Kepler} Input Catalog
(KIC; \citealp{Brown11}). The PDCSAP results are produced by the
standard \textit{Kepler} pipeline, which removes 
instrumental artifacts, and span 2009 June to 2013 May.

The sampling step of 58.9~s corresponds to $\sim4\times10^{-4}$ of the
1.7-d orbital period.  The maximum difference between `instantaneous'
and exposure-integrated model fluxes in the parameter space of
interest is 6~parts per million (ppm), which is small enough to
be neglected (deviations exceed 1~ppm for a phase range of <0.001).  

The system shows out-of-transit orbital variations arising
from \textsc{beer} effects ($\S$\ref{sec:koi}).  Even over the limited
phase range that we model, $\pm{0.1}$\porb\ around conjunction, the
amplitude of these effects is $\sim$40~ppm, which is far from
negligible.  We treated these effects as a perturbation on the basic
model, and corrected for them by using the empirical three-harmonics
model\footnote{The model defined by eqtn.~(11) and Table~5 
of \citeauthor{Shporer14} has to be reversed in both $x$ and
$y$.} described by \citet{Shporer14}.

The 25.43-hr signal has a semi-amplitude variously reported as
12--30~ppm (\citealt{Shporer11,Mazeh12,Szabo12}); in the limited
out-of-transit phase range of our data we find a semi-amplitude of
only 6~ppm, suggesting that the amplitude may be variable.  Although the period is close to a 3:5 resonance with the
orbital period \citep{Shporer11}, the ratio is not exact.
Consequently this signal is `mixed out' over the $\sim$4-year span of
the observations when phased on transits, and effectively becomes only
a minor source of additional stochastic noise.

In order to reduce the 299\,423 individual observations down to a
manageable subset for MCMC modelling, for each of 577 separate transits
the data were first phased (according to the ephemeris used in the current
MCMC cycle); corrected for \textsc{beer} effects; and rescaled to give
a median out-of-transit flux of one.\footnote{`Out of
transit' was taken as $0.045 \le |\phi| \le 0.1$, where orbital phase
$\phi$ is measured in the range $-0.5:+0.5$ about conjunction.}  In
principle, any free parameters in the adopted functional form for the
ephemeris could be allowed to `float' in the fitting process;  in
practice, we adopted a linear ephemeris with a fixed period
($\porb = 1.76358799$~d; \citealt{Shporer14}), but
allowed the time of conjunction to vary.

We then compressed the resulting data by taking median normalized
fluxes in phase bins of 0.0002 (about half the integration time of
individual observations), whence each bin contained $\sim$300 data
points.  The maximum change in normalized flux between the central
times of bins is $1.2 \times 10^{-4}$, which is comparable to the
dispersion of the individual data points ($\sim{1.6}\times{10}^{-4}$
out of transit), but large compared to the precision of the binned
data ($\sim{1.0}\times{10}^{-5}$); consequently, we tagged the median
flux in each bin with the mean time of all observations in that bin
(invariably close to the mid-bin time) rather than its original,
individual phase.

\section{Fit Results}
\label{sec:fit}

\begin{figure}
\includegraphics[width=0.47\textwidth]{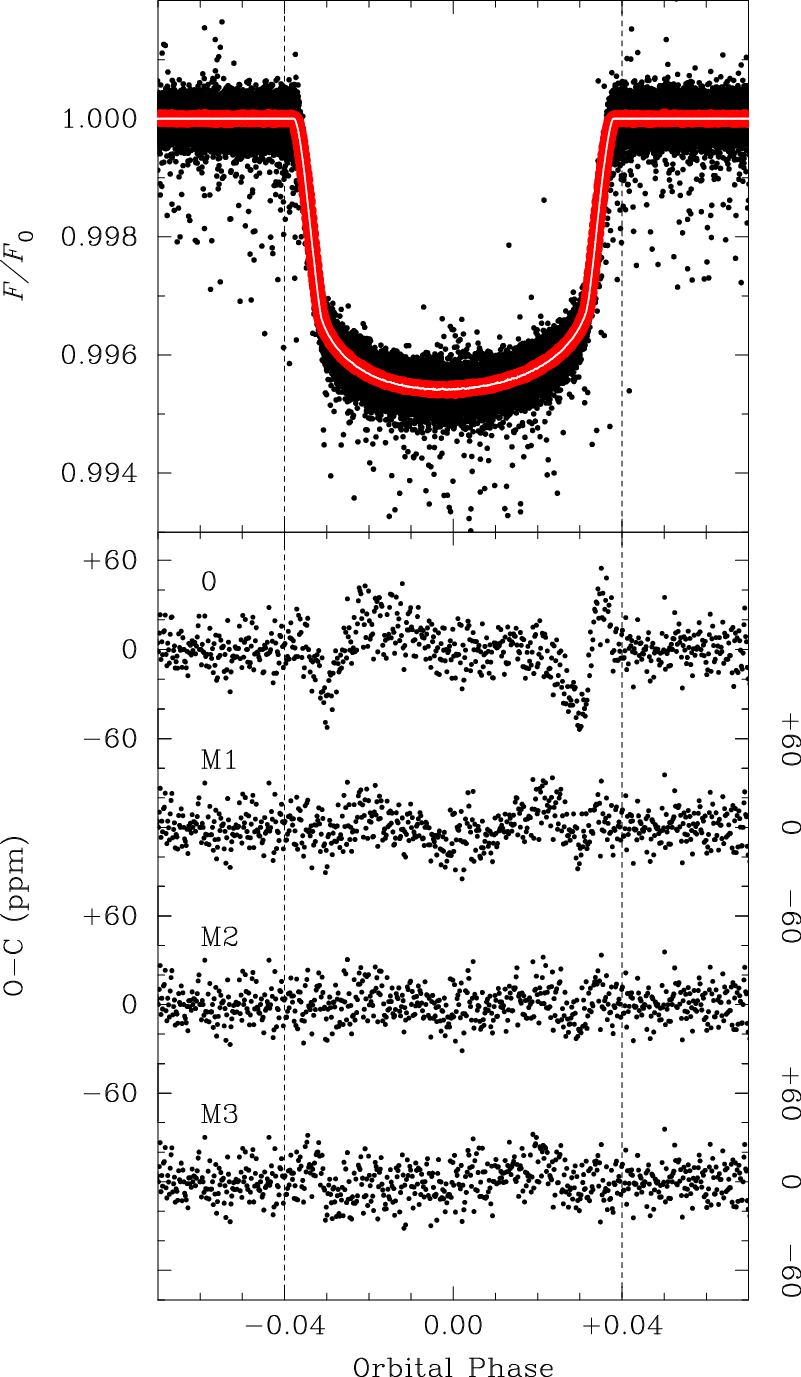}
\caption[jpeg]{Phase-folded \textit{Kepler} photo\-metry.
In the top panel, the small black dots represent individual
observations, and large red dots (which blend into a continuous band)
are the median values in phase bins of 0.002.   The white line through
the medians is from model~M2 ($\S$\ref{sec:fit});  any other gravity-darkened model is virtually
indistinguishable at the scale of this plot.\newline
The lower panel shows O$-$C residuals for different models  (cf.~$\S$\ref{sec:fit}).
Model~0 is for $\teff = 7650$~K, $L_3 = 0.45$; model~M1 is as
model~0 but with \teffL\ free;
model~M2 is as model~M1, but with \loggpL\ also free;
model~M3 is as model~M2, but with the rotation period fixed.   Vertical
dashed lines are intended simply as a visual aid to identifying transit phases.}
\label{fig:LC}
\end{figure}

As a basis for subsequent discussion, we first present the results of
an initial `maximally constrained' model, in which only (effectively)
geometric parameters were adjusted.  ELR gravity darkening
\citep{Espinosa11} and model-atmosphere
limb-darkening were used, along with fixed values for \teff\ (7650~K;
\citealt{Shporer14}) and $L_3$ (0.45; \citealt{Szabo11}).  
The results of this `model~0' are illustrated in Fig.~\ref{fig:LC}, and
show relatively large residuals during ingress
and egress ($\sim$50 ppm).

We investigated the origin of these residuals through extensive
exploration of model parameters.  Adopting eqtn.~\eqref{eq:gdark}
with $\beta$ free essentially reproduced von~Zeipel's law, which in
turn gives sensibly identical results to the ELR model
(unsurprisingly, since the latter is known to reproduce von~Zeipel at
low to moderate rotation).  Moderate adjustments to $L_3$ had
similarly small consequences for the quality of the model fits.  These
experiments identified errors in the limb darkening as the principal
cause of the discrepancies.

We addressed this issue in three ways.  First, we replaced the near-exact
represention of the angular dependence of the model-atmosphere
intensities afforded by eqtn.~\eqref{eq:cl4} with a simple quadratic
limb-darkening law,
\begin{align}
I(\mu)/I(1) = 1 - u_1(1-\mu) -u_2(1-\mu)^2,
\end{align}
with the coefficients $u_1$, $u_2$ as free parameters.  In applying
this law globally (in common with, e.g., \citealt{Masuda15}), we
abandon any latitudinal temperature dependence of the coefficients.

Secondly, in a gesture towards retaining
temperature-dependent limb-darkening while introducing only a single
additional free parameter, we investigated scaling the
linear ($a_2$) term in the
4-coefficient characterization.\footnote{There is a minor
inconsistency in
both the first and second approaches, in that the integral of intensity
over angle
will, in general, no longer \textit{exactly} match the
model-atmosphere flux, but this is unimportant for our application.}

Thirdly, recognizing that there is a temperature dependence of the
model limb darkening, we allowed the effective-temperature parameter
to float; that is, we characterize the model-atmosphere intensities
by \teffL\ rather than \teff\ ($\S$\ref{sec:teffl}).

\begin{figure}
\includegraphics[width=0.47\textwidth]{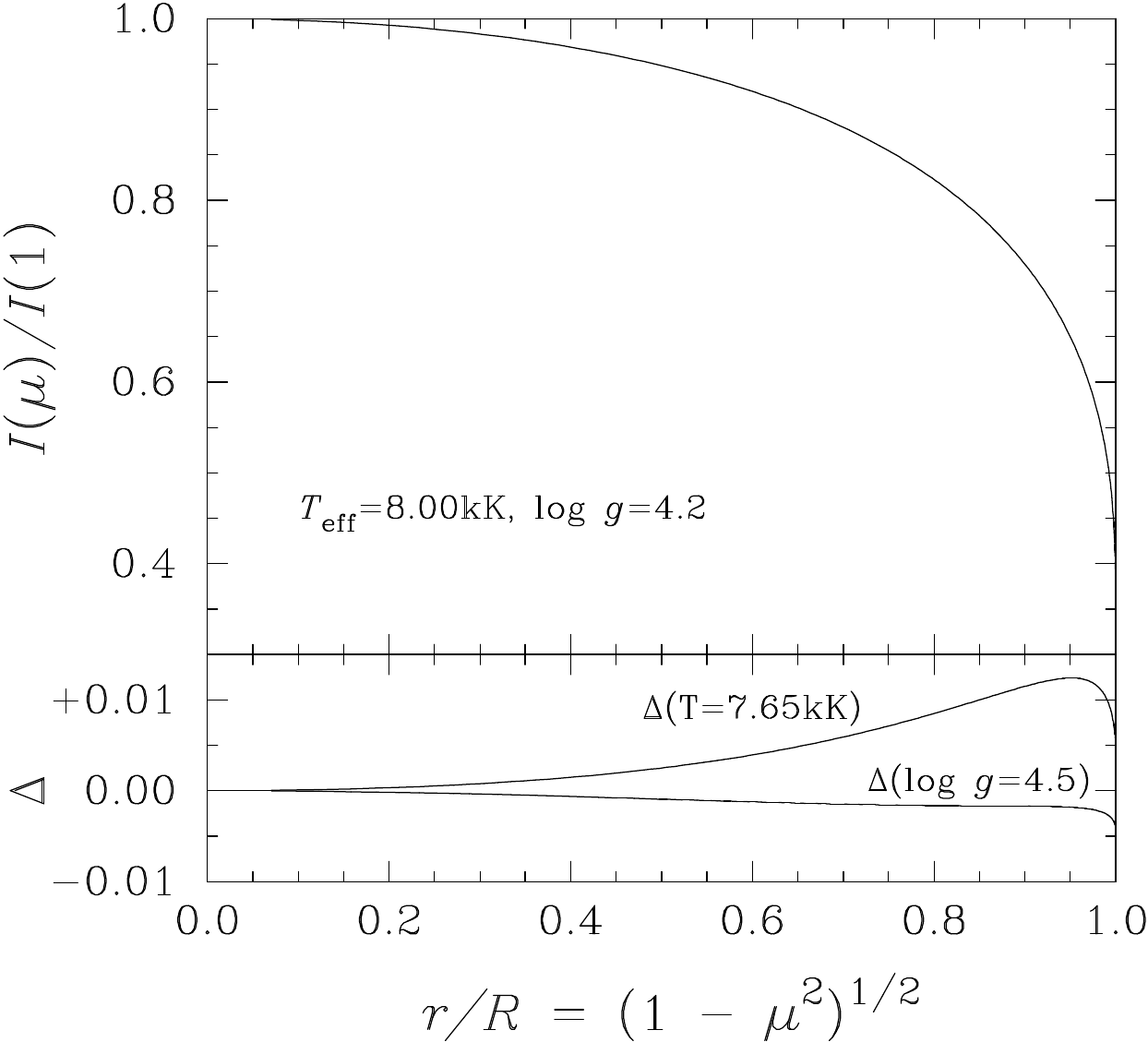}
\caption[jpeg]{Upper panel:  normalised model-atmosphere limb 
darkening at $\teff = 8.0$~kK, $\logg = 4.2$, close to values
for our best-fit models (which take into account the latitude
dependence of these parameters).   Lower panel: differences in limb
darkening for adjusted values, as indicated (in the sense reference
minus adjusted;  note the 10-fold change in
$y$-axis scale).}
\label{fig:LDC}
\end{figure}

\begin{figure}
\includegraphics[width=0.47\textwidth]{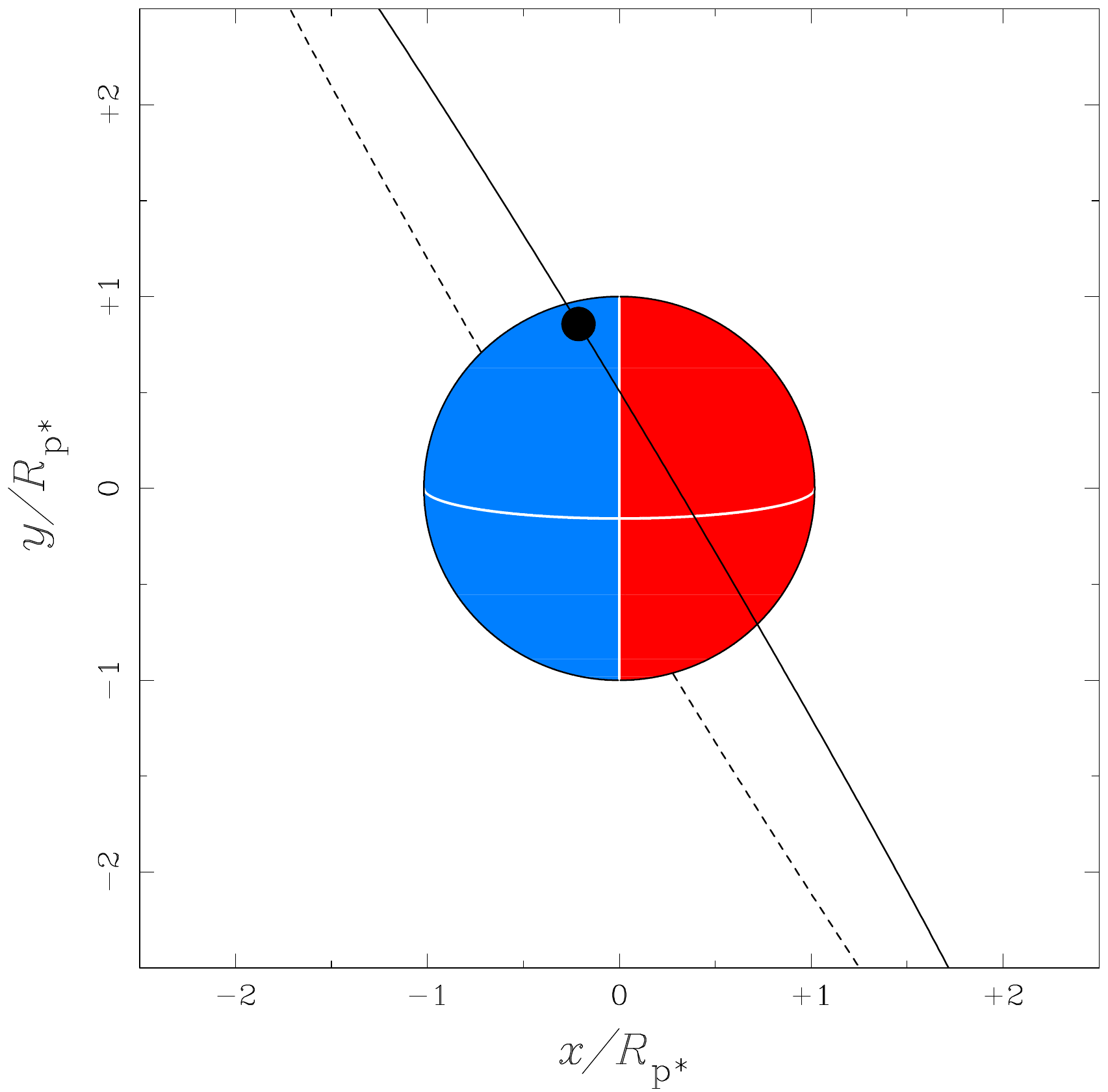}
\caption[]{Cartoon view of the system. The origin of the 
co-ordinates is the stellar centre of mass, and
the projected stellar-rotation axis
is
arbitrarily orientated along the $y$ axis; the
exoplanet orbit extends to $a \simeq 4.5 \rpole$.  The approaching and
receding stellar hemispheres are colour-coded blue and red (in the
on-line version); note that the star is \textit{slightly} oblate. 
The exoplanet is shown at orbital phase $-0.03$ (thereby indicating the
direction of orbital motion).
The model is degenerate with its mirror image about the $y$ axis.}
\label{fig:view}
\end{figure}

Unsurprisingly, all three approaches gave improved model fits, but it
is noteworthy that quite small adjustments to the model effective
temperature have significant consequences at the $\sim$10~ppm level of
precision, solely through the modest sensitivity of $I(\mu)/I(1)$ to
this parameter.  In practice, allowing \teffL\ to float also led to
smaller residuals than the other approaches in our numerical
experiments; we adopt the corresponding results for this reason, and
to avoid introducing additional ad hoc parameters.  Numerical values
for this `model~M1' are included in Table~\ref{params_tab}, and it is
confronted with the observations in Fig.~\ref{fig:LC}.
Fig.~\ref{fig:view} is a simple cartoon illustrating the implied geometry
of the system.

Model-atmosphere intensities are a function of not only temperature,
but also surface gravity (as well as abundances and microturbulence).
The true polar gravity, \loggp\ (which, with \omomc, characterizes the
overall surface-gravity distribution) is not a free parameter in our
model ($\S$\ref{sec:syspar}).  However, we can allow the value used in
obtaining
the model-atmosphere intensities, \loggpL, to `float' as,
effectively, an additional limb-darkening parameter.  Doing this
naturally affords further, albeit slight, improvement in the model fit
(model~M2 in Table~\ref{params_tab} and Fig.~~\ref{fig:LC}).

The remaining systematic residuals (peaking at
$<$10~ppm) may arise from orbital evolution over the duration
of the \textit{Kepler} observations \citep{Szabo12, Szabo14, Masuda15}, since
the time-averaged light-curve will not correspond to any single-epoch
photo\-metry.  Modelling the time-dependent behaviour is beyond the
scope of the current paper, partly because of the substantial
computing requirements required to model necessarily less compacted
datasets (we may return to this in future work), but also because our
discussion of third light ($\S$\ref{sec:l3})
emphasizes that the uncertainties on
fundamental parameters (our main interest here) are
likely to be dominated by other factors.

\begin{figure*}
\includegraphics[width=0.97\textwidth]{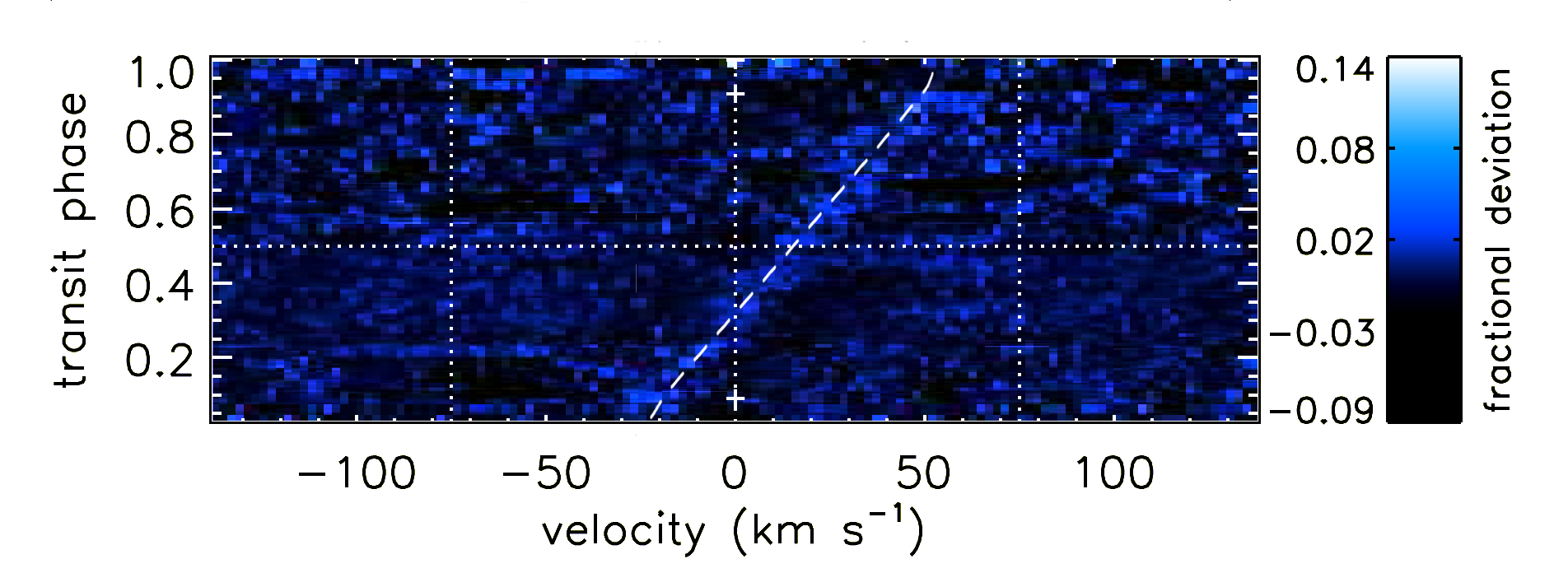}
\caption[jpeg]{Tomographic transit map, from \citeauthor{Johnson14}
(\citeyear{Johnson14}, slightly contrast enhanced), overlaid with the
prediction of the light-curve model (dashed line).   To make the 
comparison we assume that the \citeauthor{Johnson14}
`transit phase' runs from first to fourth contact, and adopt 
their value of 76.6~\kms\  for \vesini\ (which directly determines the $x$-axis scaling). }
\label{fig:tomo}
\end{figure*}

\subsection{Effective temperature and limb darkening}
\label{sec:teff}

We recall that the effective-temperature `determination' in the
model is not a traditional, direct measurement of the actual stellar
effective temperature, \teff; rather, \teffL\ is simply a parameter
which optimises model-atmosphere limb darkening (over the \textit{range} of
surface temperatures) to give a best match to the transit
data.\footnote{The same caveat
applies to \loggpL; the actual value of \loggp\ is fixed by other model parameters;
($\S$\ref{sec:syspar}).}
Only if the calculated model-atmosphere intensies are
sufficiently accurate will \teffL\ correspond to the actual effective
temperature.

However, it is noteworthy that, in practice, the optimised value
of \teffL\ falls
well within the range of direct $T_{\rm eff}$ determinations; 
while adopting only a
moderately different fixed value gives relatively large residuals.
This highlights the importance of establishing the correct value
of \teff\ when comparing empirical and theoretical limb-darkening
coefficients (or when adopting the latter).  Figure~\ref{fig:LDC}
shows the limb darkening for a model atmosphere at $\teff = 8.00$~kK,
$\logg = 4.2$, representative of the parameter space within which our
solutions fall.  The maximum difference in normalized intensity,
$I(\mu)/I(1)$, between this model and one at 7.65~kK is less than
2\%, and yet this difference accounts for almost all of the residuals for
Model~0 shown in Fig.~\ref{fig:LC}.

\subsection{Third light}
\label{sec:l3}

The third light of the unresolved optical companion KOI-13B is (literally) a nuisance
parameter in our modelling.  For our MCMC runs we experimented with
initial values of $L_3 = 0.41$--0.49 ($\Delta{m} \simeq 0.40$--0.04),
which bracket most observational determinations in the
literature\footnote{\citet{Howell11} report notably discordant values of
$\Delta{m} \simeq 0.8$--1.1 at $\sim$600--700~nm.
Although the literature values are for diverse
wavebands, the KOI-13A and~B components are of similar spectral
types and colours \citep{Szabo11}, so any wavelength dependence of
$L_3$ should be small in the optical regime.}
\citep{Fabricius02, Adams12, Law14, Shporer14}, at steps of 0.02.
%
%$\Delta{m} = 0.19 \pm 0.06$ (broad passband around $\sim$750~nm),
%\citep{Law14}, \newline
%$L_3 = 0.456 \pm .014$;\newline
%$\Delta{m} \simeq 0.0$ \citep{Adams12}\newline
%$L_3 \simeq 0.50$\newline
%Tycho double-star catalogue \citep{Fabricius02}
%$\Delta{B_T} = 0.09$
%$\Delta{V_T} = 0.13$\newline
%$L_3 = 0.48, 0.47$\newline
%\citet{Shporer14} infer a `dilution factor' 
%($=(1-L_3)^{-1}$)
%corresponding to \newline $L_3 = 0.48 \pm 0.05$.  

We found that the adopted third light always clung very close to the
initial estimate in our MCMC modelling, rather than converging onto a
value representing the global minimum in $\chi^2$ hyperspace.  This
contrasts with the behaviour of other parameters, whose values
freely migrated over relatively large ranges during `burn-in'.
Adjusting the proposal distribution did not alleviate this issue.

We believe that this outcome may arise because the transit light-curve
contains almost no information on the extent of third-light dilution
(cf., e.g., Fig.~8 of \citealt{Seager03}).  Although we might
anticipate that this should be reflected in a \textit{wide}
distribution in acceptable $L_3$ values, rather than a narrow one, in
practice the set of other parameters essentially locks in $L_3$, which
can therefore be regarded, in a limited sense, as a `derived'
parameter, given the system geometry, rather than a free one.
 
The inferred numerical values for other parameters therefore depend
somewhat on $L_3$, to a degree that typically  exceeds the formal errors on any
given model. For example, smaller $L_3$ means a shallower true transit
depth, and hence implies smaller $R_{\rm P}/R_*$ 
($\Delta(R_{\rm P}/R_*)
\simeq 0.08\Delta{L_3}$).
In recognition of this, 
while we adopt solutions with input $L_3 = 0.45$
(which yield the smallest residuals),
we give errors  in Table~\ref{params_tab} which are the quadratic sum of the
95\%-percentile ranges on those models and the
 maximum differences with the
`best-fit' parameters from models with $L_3\text{(init.)} = 0.41$--0.49
(where the latter term dominates).

\subsection{Rotation period}
\label{sec:rotper}

Our initial solutions (e.g., models~M1 and~M2) yielded rotation
periods close to 24~hr,
only $\sim$5\%\ from the 25.43-hr period found in the
\textit{Kepler} photometry
\citep{Shporer11,Mazeh12,Szabo12}.  
Although rotational modulation had not been widely anticipated for
stars hotter than the `granulation boundary' marking the transition
from radiative to convective envelopes (e.g., \citealt{Gray89}), evidence is
beginning to accumulate for starspots, of some nature, in A-type stars
\citep{Balona11, Balona17, Bohm15}, encouraging consideration of the
possibility that we are seeing a rotational signature in
KOI-13 
($\teff \simeq 8$~kK corresponds to spectral type A5--A7), as
suggested by \citet{Szabo12}.

We can impose the constraint of fixed \prot\ on the model,
which links \omomc\ to $\rpole/a$ in the MCMC chains
(Appendix~\ref{appx:one}, eqtn.~\ref{eq:app2}).  The results of this
model~M3 are reported in Table~\ref{params_tab}; the fit quality is quite reasonable
(Fig.~\ref{fig:LC}).
Because the transit
depth essentially fixes $R_{\rm P}/R_*$, the main effect of imposing
a longer rotation period is to decrease the angular rotation rate,
which for given \vesini\ leads to a larger stellar radius, and hence,
for $\sim$fixed density, a higher stellar mass, as discussed in the
Section~\ref{sec:syspar}.

\subsection{Tomo\-graphy}

There are no published Rossiter--McLaughlin investigations of KOI-13, but
\citet{Johnson14} conducted a detailed tomo\-graphic study,
providing a velocity-resolved map of the transit.

Our model allows stellar velocities (R--M effect or tomo\-graphic
counterpart) to be evaluated directly.  This can be accomplished by
synthesizing the spectrum as a function of orbital phase, and
subjecting the ensemble of synthetic spectra to the same analysis as
the observations (e.g., cross-correlation, or tomo\-graphy).  However, for
the present study we simply take the intensity-weighted average radial
velocity,
\begin{align*}
v(\lambda) = \frac{\int{v \times
I(\lambda, \mu, \teffl, g)\,{\rm d}A}} {\int{
I(\lambda, \mu, \teffl, g)\,{\rm d}A}}
\end{align*}
where the integration is over area, and the (weak) wavelength
dependence of the model velocity comes about because of the 
wavelength dependence of intensities on
limb darkening and temperature.  To evaluate the R--M
effect the integration is conducted over all visible elements, while
taking the velocity of all occulted elements models the tomo\-graphic
signature.

The predicted locus of velocity vs.\ phase from the light-curve
solution is compared to the \citeauthor{Johnson14} map in
Fig.~\ref{fig:tomo}.  The agreement is very
satisfactory, arising from the accord between the values of projected
obliquity $\lambda$ and impact parameter $b$ obtained from the
\textit{independent} tomo\-graphic and photometric solutions
($\Delta\lambda = 0{\fdg}6 \pm 2{\fdg}0$, \mbox{$\Delta{b} = 0.01\pm 0.03$}).

\section{System parameters}
\label{sec:syspar}

Any fundamentally geometric transit model, such as employed here, is
of necessity scale free.  Consider Fig.~\ref{fig:view}; there is no
indication of whether this is a small, nearby system, or a large,
distant one.

Nevertheless, for given orbital period, a large, distant system must have
greater orbital velocities, and hence greater masses, than a smaller,
nearby system.  This relationship between scale and mass is codified
in Kepler's third law, which leads directly to a constraint on
$a^3/(\mstar+M_{\rm P})$, and hence, given the dimensionless radius
$\rstar/a$, to the
stellar density (e.g., \citealt{Seager03}) -- but not the mass and radius separately.

\citet{Barnes11} suggested that rotational
effects, and specifically gravity darkening, can, in principle, lift
the ``density degeneracy'', through the dependence of $\Omega$ on mean
stellar radius $R_*$.  However, in the Roche approximation the light-curve
depends on rotational effects only through the ratio $\omomc$; to get to $\Omega$ requires
calculation of $\Omega_{\rm c}$, which itself has an $M/R^3$
dependence.  Consequently, $\Omega$ is actually scale-free (as shown
analytically in Appendix~\ref{appx:one}), and a Roche-model analysis of
the transit light-curve alone cannot break the mass/radius
degeneracy in $M/R^3$.

Of course, if the orbital velocities can be established for both
components, these determine the absolute scale -- the standard
`double-lined eclipsing binary' approach.  However, an alternative,
independent means of establishing the orbital semimajor axis (and
hence other system parameters) is available if \prot, the stellar
rotation period, \istar, the axial inclination, and \vesini, the
line-of-sight component of the equatorial rotation speed, can be
determined; these immediately yield the equatorial radius,
\begin{align*}
\reqtr = (\prot \vesini)/(2 \pi \sin{i})
\end{align*}

The quantities
\prot\ and \istar\ can be estimated if the circular symmetry
of the projected stellar disk is broken.  A familiar example is when
starspots are present, but gravity-darkened stars have the same
potential (since \omomc\ relates, indirectly, to \prot).  Introducing
the observed projected equatorial rotation speed, \vesini, as a
constraint on the light-curve solution therefore affords usefully
tight limits on the absolute dimensions of the system.  The
straightforward algebra is set out in Appendix~\ref{appx:one}.

There are two precise determinations of projected rotation speed of
KOI-13A in the literature, in good mutual agreement: $\vesini =
76.96 \pm 0.61$~\kms\ and $76.6 \pm
0.2$~\kms\ \citep{Johnson14,Santerne12}.  We adopt the latter, more
precise value
in order to calculate the system dimensions
reported in Table~\ref{params_tab}.

[Our referee raised the point that the precision of these results may
not reflect their accuracy, an observation with which we fully concur
(cf., e.g., \citealt{Howarth04}).  However, as shown in
Appendix~\ref{appx:one} (eqtn.~\ref{eq:app2}), the semi-major axis
scales linearly with \vesini; radii converted from normalized to
absolute values scale in the same way, while the absolute system mass
scales as $(\vesini)^3$, from Kepler's third law.  Hence the results,
or uncertainties, are readily reassessed if another value for the
projected equatorial rotation speed is preferred.]

%\subsection{Comparison to models}
%
%We compared the results to tailored stellar-evolution models
%calculated with \textsc{mesa} \citep{Paxton11,Paxton13,Paxton15} through the \texttt{MESA-Web} interface
%\citep{Fields15}.

%\subsection{Comparison with previous work}
%
%Although the transit light-curve is scale-free,  
%\citet{Barnes11} and \citet{Masuda15} chose to impose 
%fixed stellar masses in their model realizations, 
%in order to obtain
%physical dimensions directly for other parameters.
%
%Masuda fixes \mstar, and fits for $\rho_*$, which he defines as
%$3\mstar/4\pi\reqtr^3$, and for the rotational frequency, which acts
%as a surrogate for the oblateness.  The mass and orbital period give
%the (assumed) orbital semi-major axis, while the mass, density, and
%oblateness give \rpole.
%
%
%The Barnes \& Masuda models don't work so well for this (because an
%oblate spheroid gives different answers to a Roche model for veq).

\subsection{Distance}

The effective temperature determines the 
surface brightness;  
given the size of the star the
absolute magnitude follows, and hence the
distance.   We find
\begin{align*}
M(V) \simeq 2.44 + 0.51\left({ 8.0 - \frac{\teff}{\text{kK}} }\right)
- 5\log\left({  \frac{\rpole}{1.49\rsun}}\right)
\end{align*}
where the second term is an empirical fit to models with $7.5
< \teff/\text{kK} < 8.5$; model-atmosphere Johnson $V$-band fluxes are
from \citet{Howarth11a}; and we neglect the further, unimportant,
dependences of $M(V)$ on \omomc\ and \istar.

There is a surprisingly large dispersion in the photometry of KOI-13 catalogued in
the \textit{Vizier} system of the 
Centre de Donn\'ees astronomiques de Strasbourg, most of which clearly
refers to the combined light of the visual binary.   
%
%   9.1     SAO cat
%  10.00    Tycho input catalogue
%   9.95    Hipparcos & Tycho catalogues
%  10.455   NOMAD Catalog (Zacharias+ 2005) 
%  10.35    The Guide Star Catalog, Version 2.3.2
%   9.697   UCAC4 Catalogue & All-sky catalog of solar-type dwarfs 
%   9.734   TASS Mark IV patches photometric catalog, version 2 
%   9.950   SKY2000 Catalog, Version 4, Version 5 (same)
%   
%   9.9  +10.2     The Catalogue of Components of Double and Multiple Stars
%                  -> 9.3
%  10.109+10.446   All-sky Compiled Catalogue of 2.5 million stars
%                  -> 9.512
%  10.326+10.455   Transformed 	The Tycho-2 Catalogue (Hog+ 2000) 
%                  -> 9.636
We adopt the spatially resolved Tycho-2
photometry, which transforms to $V = 10.33$ for \mbox{KOI-13A} (with
an uncertainty of $\sim$0.05; \citealt{Hog00}).  Foreground reddening
is estimated as $E(B-V) \simeq 0{\fm}02$ from \citet{Green15},
 whence
\begin{align*}
\log\left({ \frac{d}{\text{pc}} }\right) &=
2.566 + 0.2\left[{
(V - 10.33) - (A(V)  - 0.06)
}\right]
\\
&\quad -0.102\left({ 8.0 - \frac{\teff}{\text{kK}} }\right)
+ \log\left({  \frac{\rpole}{1.49\rsun}}\right);
\end{align*}
i.e., $d\simeq 370$~pc, with an uncertainty of perhaps $\sim$25~pc.
% The only other distance estimate in the literature is
% $d = 530$~pc, reported by 
% \citet{Pickles10}; this is, essentially, a spectroscopic parallax from
% a photometry-based spectral classification.

\section{Conclusions}

We have conducted a new solution of \textit{Kepler} photometry of
transits of KOI-13b, obtaining results that are substantially in
agreement with those found by \citet{Masuda15}, and in accord with the
tomo\-graphy reported by \citet{Johnson14}.  The solution yields both
the projected and true angular separations of the orbital and
stellar-rotation angular-momentum vectors.  We emphasize that any
photometric solution is necessarily scale-free (e.g., does not require
a stellar mass to be assumed); but demonstrate that,
by adopting a value for \vesini, the absolute system dimensions and
mass can be established.  Allowing for the full range of solutions
(Table~\ref{params_tab}; third light $L_3 = 0.41$--0.49, free or fixed
stellar rotation period), we obtain a planetary radius
$R_{\rm P}/\rjup = 1.33 \pm 0.05$,
stellar polar radius 
$\rpole/\rsun = 1.55 \pm 0.06$, and a
combined mass 
$\mstar + M_{\rm P} (\simeq{\mstar}) = 1.47 \pm 0.17$~\msun.
All solutions place KOI-13 in an unremarkable location in the
main-sequence mass--radius plane (e.g., \citealt{Eker15}).

% In this sense, the 
% model \vesini, and the amplitude
% (but not the shape) of the Rossiter--McLaughlin effect, are
% arbitrary.   The stellar radius (and distance) can be inferred from an
% observed \vesini.
%
% 
% Adjust g(pole) at fixed R(pole) and Om/Om(crit) 
% to get P(orb) right -- doesn't affect geometry.
% 
% Then adjust Rpole to get v.sin(i) right.  But changing *only* R(pole)
% changes P(orb) (by changing the linear scale of the system), so we
% change $M/R^3 == g/R$.  This retains geometry *and* P(orb).
% 
% Result:
% g(pole) = 3.9615 (M*=2.348)
% R(pole) = 2.6515
% 
% Original g(pole) was 4.2 -- negligible differences to l-c.
% [really?  Significant changes on re-examination!]

%estimating $\teff \simeq 8.2\pm 0.4$~kK, and

%$\Delta{m} = 0.19 \pm 0.06$ (broad passband around $\sim$750~nm;
%\citealt{Law14}) \newline
%$L_3 = 0.456 \pm .014$
%
%$\Delta{m} = 0.8$--1.2 ($\sim 600$~nm)
%\citealt{Howell11}
%
%$\Delta{m} \simeq 0.0$ \citep{Adams12}\newline
%$L_3 \simeq 0.50$
%
%Tycho double-star catalogue \citep{Fabricius02}
%$\Delta{B_T} = 0.09$
%$\Delta{V_T} = 0.13$\newline
%$L_3 = 0.48, 0.47$
%
%
%\citet{Shporer14} infer a `dilution factor' 
%%($=(1-L_3)^{-1}$)
%corresponding to \newline
%$L_3 = 0.48 \pm 0.05$
%
%
%So L3 0.456 0.48 0.47 0.48
%

%Rotation $\vesini \simeq 65$--70~\kms\ \citep{Szabo11}

%Adopt 76.6

% $\lambda = 58{\fdg}6 \pm 2{\fdg}0$,
% $b  = 0.256 \pm 0.030$ \citep{Johnson14}.

%$\logg = 3.9 \pm 0.1$ \citep{Szabo11}
%
%$P_orb = 1.76358799$ \citep{Shporer14}

\bibliographystyle{mnras}
\bibliography{IDH}

\appendix
\section{ Scaling}
\label{appx:one}

The photo\-metric solution establishes reasonably precise values for
$\rpole/a$, \omomc, and $\sin{\istar}$; and we have the ancillary
observational quantities
\porb\ and \vesini\ to good accuracy.  

In the Roche model,
the critical angular rotation rate at which the equatorial surface
gravity is zero is
\begin{align*}
\Omega_{\rm c} = \sqrt{ \frac{8}{27} \frac{G \mstar}{\rpole^3} }.
\end{align*}
%(e.g., \citealt{Howarth01}).  
The equatorial rotation speed is
\begin{align}
v_{\rm e} &= \Omega \reqtr = \left({\omomc}\right) \Omega_{\rm  c} \,
f \rpole\nonumber \\
&= f \left({\OmOmc}\right) \sqrt{ \frac{8}{27}\frac{G\mstar}{\rpole} }  
\label{eq:veq}
\end{align}
where, again in the Roche model, the function $f$ is given by
\begin{align*}
f = \frac{\reqtr}{\rpole}
= \frac{3}{(\omomc)} 
\cos\left[{
\frac{\pi + \cos^{-1}(\omomc)}{3} }\right]
\end{align*}
\citep{Harrington68}.    Using Kepler's third law,
\begin{align*}
\mstar + M_{\rm P} &\equiv \mstar(1+q)\nonumber \\
&= \frac{4\pi^2}{G\,\porb^2}\frac{\rpole^3}{(\rpole/a)^3},
\end{align*}
for the mass in eqtn.~\eqref{eq:veq}, and rearranging, gives
the semi-major axis:
\begin{align}
a =
\frac {\porb}{f\,(\omomc)}
\frac {\vesini}{\sin{\istar} }
\sqrt{ \left({\frac{\rpole}{a}}\right)
\frac{27\,(1+q)}{32\pi^2} }.
\label{eq:app2}
\end{align}
All terms on the right-hand side are `known', except the mass ratio $q
= M_{\rm P}/\mstar$, which it 
may often be reasonable to assume to be negligibly
small if no numerical estimate is available.
Having evaluated the orbital semi-major axis, the linear
dimensions of the system components, and the mass, follow
(radii from $R/a$, and \mstar\ from Kepler's third law).

Using similar reasoning as above, we also have
\begin{align}
\prot &= \frac{2\pi}{\Omega} \nonumber\\
      &= \frac{2\pi}{(\omomc)} \sqrt{ \frac{27}{8} \frac{\rpole^3}{G\mstar} }\nonumber\\
      &= \frac{\porb}{(\omomc)} \sqrt{ \left(
      {\frac{3}{2}\frac{\rpole}{a}}\right)^3(1+q)}.
\label{eq:app3}
\end{align} 
Thus in the Roche model the rotation period (or, equivalently,
$\Omega$) is scale free, and of itself offers no independent leverage on
absolute values of \mstar\ or \rpole.   However, if \prot\ is known from
independent considerations, it may be used to constrain the
combination
$(\rpole/a)^{3/2}(\omomc)^{-1}$ ($\S$\ref{sec:rotper}).

\label{lastpage}

\end{document}